\documentclass[
amsmath,amssymb,
reprint,%
]{revtex4-2}

\usepackage{graphicx}
\graphicspath{{Figures/}}
\usepackage{floatrow}
\usepackage{makecell}
\usepackage{float}

\usepackage{dcolumn}
\usepackage{bm}
\usepackage[mathlines]{lineno}

\usepackage[utf8]{inputenc}
\usepackage[T1]{fontenc}
\usepackage{mathptmx}
\usepackage{xcolor}

\usepackage{siunitx}
\usepackage{amsfonts}

\begin{document}

\title{Magnetic Reconnection on a Klein Bottle}
	
	\author{Luke Xia}
    \email{lyxia@uci.edu}
 	\affiliation{University of California, Irvine 92697, USA}
	\author{M. Swisdak}
    \email{swisdak@umd.edu}
	\affiliation{Institute for Research in Electronics and Applied Physics, University of Maryland, College Park, Maryland 20742, USA}

	\date[]{Presented XXXXX; received XXXXX; accepted XXXXX; published online XXXXX}

	\begin{abstract}
	We present a new boundary condition for simulations of magnetic reconnection based on the topology of a Klein bottle.  When applicable, the new condition is computationally cheaper than fully periodic boundary conditions, reconnects more flux than systems with conducting boundaries, and does not require assumptions about regions external to the simulation as is necessary for open boundaries.  The new condition reproduces the expected features of reconnection, but cannot be straightforwardly applied in systems with asymmetric upstream plasmas.
	\end{abstract}

\maketitle

\section{Introduction}\label{sec:intro}

During reconnection oppositely directed components of a magnetic field come together at an X-point and undergo a topological reordering that enables the transfer of magnetic energy to the surrounding plasma.  In the collisionless plasmas of space and astrophysics, reconnection has particular importance due to its role in catalyzing energy release over a wide range of scales in flares, jets, and other phenomena.


Unsurprisingly then, numerical simulations of reconnection have been a subject of significant interest.  Any simulation entails certain fundamental choices, among them the boundary conditions imposed on the computational domain.  Idealized simulations -- i.e., those not modeling a particular event -- often begin in a configuration based on the Harris sheet \citep{harris62a}, a steady-state solution of the Vlasov equation in which the magnetic field has the simple profile 
\begin{equation}\label{eq:gem}
\mathbf{B} = B_0 \tanh(y/w_0)\,\mathbf{\hat{x}},
\end{equation}
with $w_0$ a constant.  (For definiteness, we assume the simulation domain occupies the space $-L_x/2\leq x \leq L_x/2$, $-L_y/2 \leq y \leq L_y/2$. A fully three-dimensional domain is not necessary for reconnection to occur and so the $z$ coordinate will be suppressed.)  The reversal of the field across $y=0$ implies the existence of a current sheet, with the current pointing in the $-z$ direction.

The form of the magnetic field in the Harris sheet imposes constraints on the boundary conditions.  Periodic boundaries are typically used in the horizontal ($x$) direction, but the anti-symmetric nature of $\mathbf{B}$, $B_x(y=-L_y/2) = -B_x(y=L_y/2)$, precludes the use of periodic boundaries in the vertical ($y$) direction.  Two options are usually considered.  The first is the inclusion of hard conducting walls at the top and bottom boundaries of the domain.   Appropriate boundary conditions are imposed on the electromagnetic fields (e.g., Dirichlet for $E_x$, $B_y$, and $E_z$ and von Neumann for the other components) while particles specularly reflect from the walls.  The influential GEM Challenge \citep{birn01a} used these conditions and so hereinafter we will use its name to denote them. For the GEM Challenge boundary conditions the computational domain is topologically equivalent (homeomorphic) to a cylinder.

The second common option is to add a second current sheet of opposite orientation so that
\begin{equation}\label{eq:doubleharris}
\mathbf{B} = B_0 \left[\tanh\left(\frac{y-L_y/4}{w_0}\right) - \tanh\left(\frac{y+L_y/4}{w_0}\right) +1\right] \,\mathbf{\hat{x}}
\end{equation}
This choice allows periodic boundary conditions in the vertical direction because $B_x(y=-L_y/2) = B_x(y=L_y/2)$.  (For both choices, one usually also requires $w_0\ll L_y$ so that $\partial B_x/\partial y$, and hence the current, is negligible for $y = \pm L_y/2$.)  The implementation of fully periodic boundaries is typically straightforward since no special conditions need to be placed on the electromagnetic fields or particles at any point in the grid.  In this case, the computational domain is homeomorphic to a torus.

Either choice comes with limitations.  Simulations using the GEM Challenge boundary conditions will stagnate when the reconnected flux accumulating in the magnetic islands presses against the hard boundaries (see Figure \ref{fig:jez_gem_klein} below).  Fully periodic boundary conditions employing the field described in equation \ref{eq:doubleharris} can mitigate this stagnation by offsetting the X-points on the two current layers by $L_x/2$, thus allowing the flux accumulating in the island of one layer to drive inflow towards the X-line on the other. 
However, keeping the amount of flux available to reconnect the same for both types of boundary conditions (equivalently, keeping the aspect ratios of the current sheets the same) requires a doubling of the size of the computational domain, and hence the computational cost, in the fully periodic case.  Less-common alternatives, e.g., open boundaries or driven reconnection \citep{horiuchi01a,daughton06a,divin07a,roytershteyn10a}, increase the computational complexity and can only partially mitigate these issues (often while raising new ones).

Here we present a novel type of periodic boundary condition in which the computational domain is homeomorphic to a Klein bottle. Making this choice allows the simulation to enjoy the benefits of the two major options discussed above: the computational expense of the GEM Challenge conditions coupled with the self-driving of the fully periodic domain.  We show that these boundary conditions preserve the important properties of magnetic reconnection.  Section \ref{sec:methods} discusses our computational methods, including the implementation of the new boundary condition.  Results from the simulations are presented in Section \ref{sec:results}, while Section \ref{sec:disc} offers our conclusions.

\section{Computational Methods}\label{sec:methods}
\subsection{Code}
The simulations are performed with {\tt p3d}, a massively parallel particle-in-cell (PIC) code with a long history of simulating magnetic reconnection \citep{zeiler02a}.  It employs units in which a reference density $n_0$ and magnetic field strength
$B_0$ define the units of length, the
proton inertial length $d_{p0} =c/\omega_{p0}$ (where $\omega_{p0} = \sqrt{4\pi n_0 e^2/m_p}$ is
the proton plasma frequency), and time, the proton cyclotron time
$\Omega_{p0}^{-1} = m_pc/eB_0$. 
Velocities are measured in units of a reference Alfv\'en speed
$v_{A0}=\sqrt{B_0^2/4\pi m_pn_0}$ while electric fields and temperatures are normalized
to $v_{A0}B_0/c$ and $m_pv_{A0}^2$, respectively.  The simulations presented here follow particles in a 2x3v (also known as 2.5D) phase space in which no variations are allowed in the third spatial dimension (i.e., $\partial/\partial z = 0$).  

With the exception of a slight alteration in the size of the computational domain, the parameters are chosen to match those of the GEM Challenge \citep{birn01a}.  In order to reduce the separation between the spatial scales of the protons and the electrons, and hence the computational
expense, the electron mass is taken to be $m_e/m_i = 0.04$, which
implies the electron inertial length $d_e = 0.2d_p$.  For
similar computational reasons, the ratio of the speed of light to the
Alfv\'en speed (equivalently, the ratio of the proton plasma and
cyclotron frequencies) is taken to be 20.  The initial electron and proton temperatures are $1/12$ and $5/12$, respectively.  The timestep is chosen to satisfy the CFL condition and, as a consequence, particles can travel at most one grid cell per timestep.  


For later reference, it is useful to discuss the layout of the computational domain in more detail than usual.  To parallelize a computation, {\tt p3d} breaks the domain into subdomains in configuration space, each of which is assigned to a single processor.  The processors are organized into a rectangular grid of size $P=P_x \times P_y \times P_z$ and each processor is further subdivided into $N_x \times N_y \times N_z$ grid points.  For notational simplicity we will assume $N_z=P_z = 1$ in what follows, but the extension to multiple processors and gridpoints in the third dimension is straightforward.  Each processor has an overall label that can range from $0$ to $P-1$ as well as a label in each dimension:  $p_x\in \{0, 1, 2, ..., P_x-1\}$, $p_y\in \{0, 1, 2, ..., P_y-1\}$, and $p_z\in \{0, 1, 2, ..., P_z-1\}$.  The overall label and the triplet $(p_x,p_y,p_z)$ are unique to each processor and it is straightforward, given the dimensions of the computational domain, to convert between the two.  For example, the processor at the bottom-left of the domain, $P=0$,  has triplet $(0,0,0)$.  The processor to its immediate right, $P=1$, has triplet $(1,0,0)$ and the processor immediately above is $P=P_x$ with triplet $(0,1,0)$. 
For communication purposes, every processor needs to know the identities of the other processors bordering it, and while this  calculation is simple for those on the interior of the computational domain, it can be more complicated for those on the boundaries.



\subsection{Klein Bottles and Processor Mapping}

\begin{figure}
\begin{center}
\includegraphics[width=0.5\textwidth]{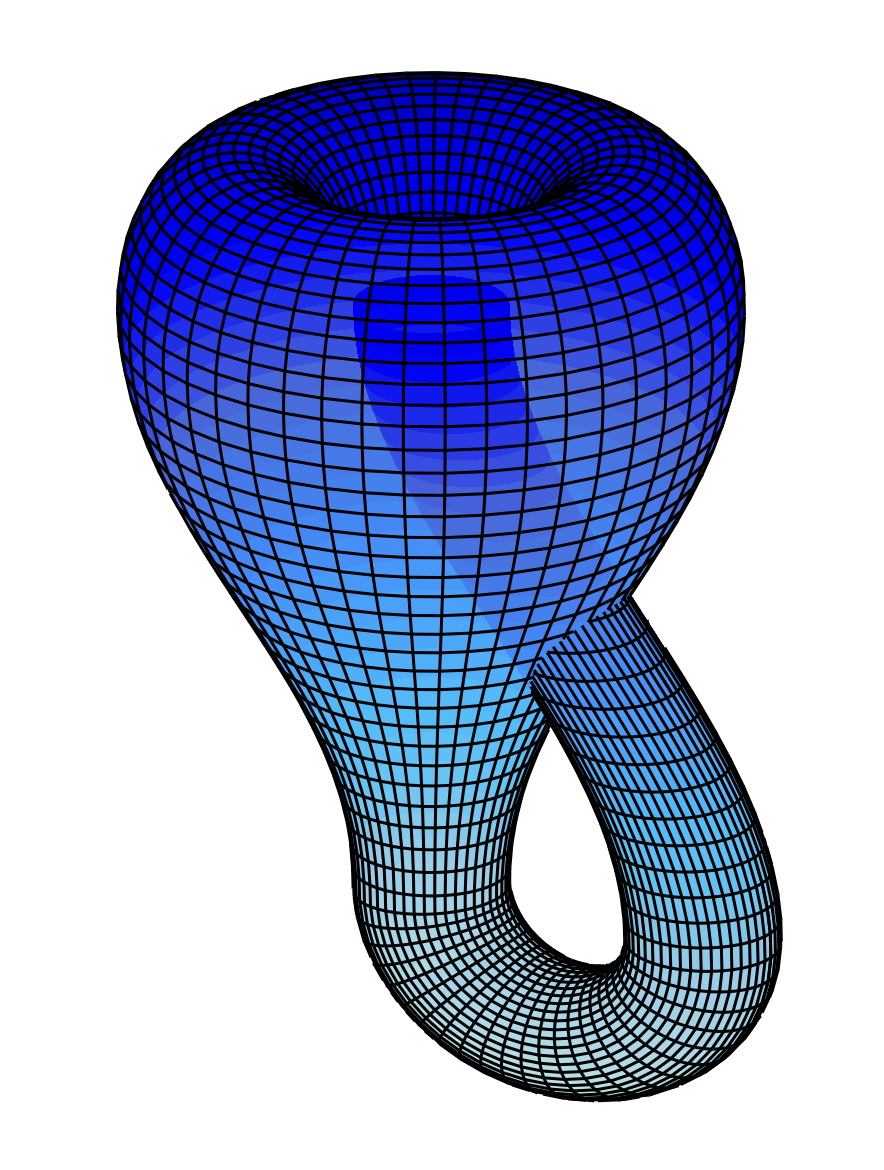}
\caption{Immersion of a two-dimensional Klein bottle in three-dimensional space.}\label{fig:kleinbottle}
\end{center}
\end{figure}

The novel boundary condition presented here is based on the topology of a Klein bottle, a two-dimensional non-orientable manifold.  (The perhaps-more-familiar M\"obius strip is another example of a non-orientable manifold.)  A Klein bottle is a surface with no distinct front or back side that can be obtained by identifying and gluing together specific edges of a rectangle in a non-trivial way, resulting in a structure that cannot be embedded smoothly in three-dimensional space without self-intersections or singularities (see Figure \ref{fig:kleinbottle} for a representation).  
Higher-dimensional analogues of Klein bottles can be similarly constructed.  For example, mapping the edges of a rectangular prism in a particular way produces
a system homeomorphic to a Klein geometry in $\mathbb{R}^4$.  






\begin{figure}
\begin{center}
\includegraphics[width=\textwidth]{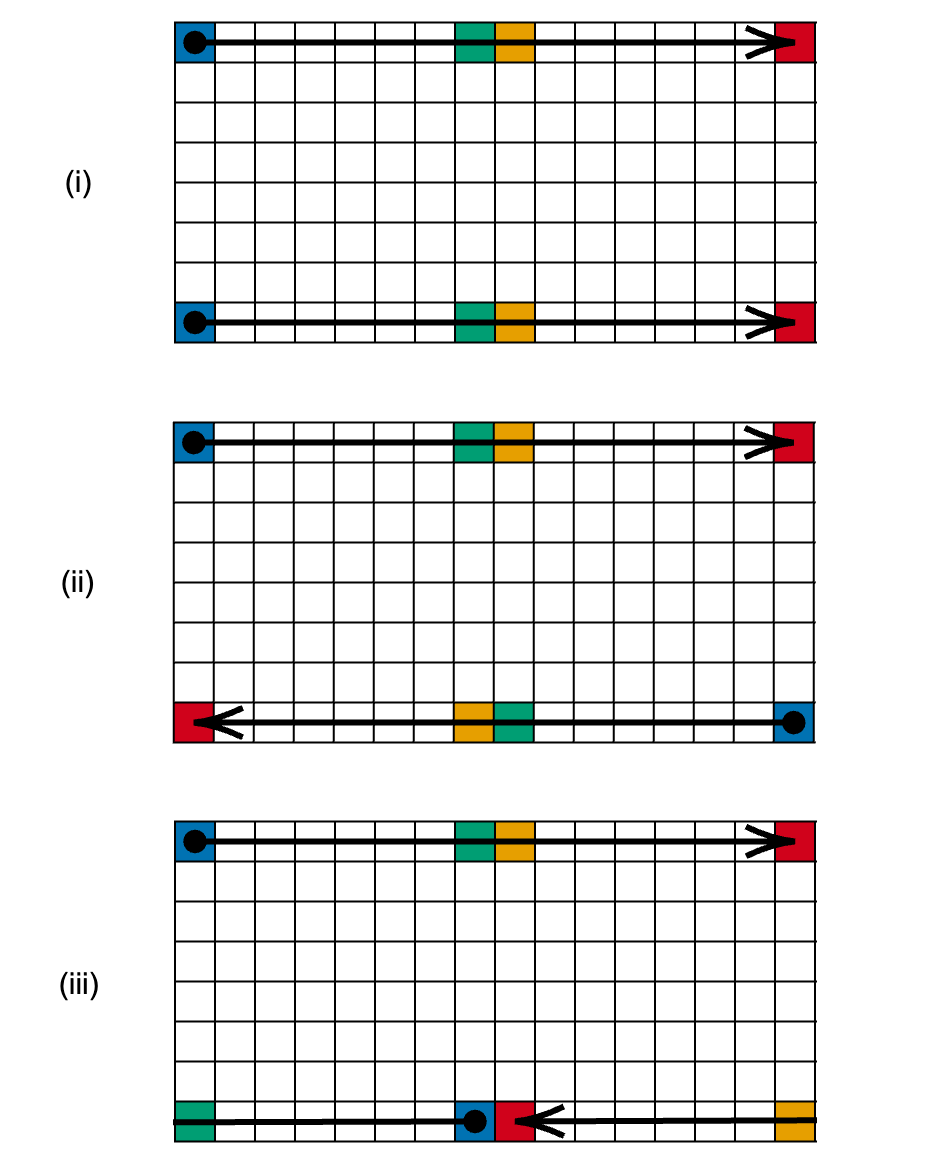}
\caption{Topological mappings of (i) a torus (ii) a Klein bottle (iii) a phase-shifted Klein bottle.  The horizontal boundaries are periodic.  Squares represent processors and those of the same color abut in the vertical direction.   The arrow demonstrates how a vector transforms across the boundary.}\label{fig:mappings}
\end{center}
\end{figure}

In numerical simulations, the identification and gluing-together process is equivalent to describing how processors abut one another when crossing a boundary.  This process is given a pictorial representation in Figure \ref{fig:mappings}.   In this visualization, small squares represent individual processors; squares painted the same color neighbor one another in the vertical direction. (Only the top and bottom boundaries are of interest, as the horizontal direction is assumed to be periodic in each panel.)  In the configuration shown, with $P = P_x \times P_y \times P_z = 16\times8\times1 = 128$, the processors are numbered from $0$ to $127$, with those on the bottom row having $P \in\{0, 1, ... 15\}$.
The fact that the doubly periodic boundary conditions described in Section \ref{sec:intro} are topologically equivalent to a torus is demonstrated in the first panel.   Travelling downward from the bottom row brings one to the top row, where the processors have $P\in\{112, 113, ... 127\}$.  In other words, processors next to the boundary (on either side) with triplets of general form $(i,j,0)$ border processors with triplets of the form $(i,P_y-(j+1),0)$.  The overplotted arrow illustrates how a vector transforms when moving between the top and bottom rows of processors.  Note that interpreting the arrow as a magnetic field vector makes it clear why a system with a single current sheet -- across which the field must switch direction -- cannot use these boundary conditions.

To map a two-dimensional sheet to a Klein bottle, one connects the left and right sides together as usual. However, processors on the top and bottom boundaries now adjoin counterparts at mirrored positions on the $x$-axis, as shown in Figure \ref{fig:mappings}(ii). A processor on the top or bottom boundary with triplet of the form $(i,j,0)$ borders a processor with triplet $(P_x-(i+1),P_y-(j+1),0)$.  (Note the set of neighboring gridpoints within processors are also mirrored, so that the leftmost gridpoint on, for example, the bottom row of processor 1 is considered to be directly above the rightmost gridpoint in the top row of processor 126.)  As a consequence, vectors flip orientation across the boundary, as shown by the overplotted arrow.   Implementation of this boundary condition requires several algorithmic changes.  Processors on the top and bottom rows have a new set of neighboring processors and the $x$ and $z$ components of any vector quantity, e.g., particle velocities and electromagnetic fields,  must be flipped across the boundary.  (In practice, processors in {\ttfamily p3d} interact with neighbors through  guard cells.  Any changes to the signs of gridded quantities required by the boundary conditions occurs when these cells are updated.) This means that only one current sheet needs to be included in the domain, as the reversal in $B_x$ is naturally accommodated.




However, the basic Klein boundary condition shown in Figure \ref{fig:mappings}(ii) is not ideal for reconnection simulations.  The reason becomes clear after consideration of the magnetic configuration at late times.  Reconnection proceeding at the X-line (assumed to be at the center of the domain), produces an island of magnetic flux centered on the left/right (they are equivalent) boundary.  This island grows, eventually becoming large enough to approach the top and bottom boundaries.  In runs employing the GEM Challenge boundary condition, this island presses against the conducting walls, producing a back-pressure that resists further reconnection and eventually stifles the simulation.  The same issue bedevils the Klein boundary condition, although in that case the back-pressure is not provided by conducting walls but by the island itself -- growth of the island in, for instance, the bottom-left of the domain, presses against the island in the upper-right.

Fortunately, a straightforward adjustment that mimics the behavior of doubly periodic boundary conditions can alleviate this issue by, in essence, creating self-driving reconnection.  Arranging for the island growing on the left and right sides of the domain to emerge from the top and bottom boundaries in the middle of the domain extends the time that reconnection will occur.  The phase shift is produced by displacing the set of neighboring processors for either the top or bottom boundary by \(L_x/2\), forming the configuration displayed in Figure \ref{fig:mappings}(iii). Returning to the example of a $16\times8$-processor domain, each processor with $P\in\{0, 1, \dots 7\}$ now neighbors the corresponding processor $P\in\{119, 118, \dots 112\}$, and vice versa, while each processor $P\in\{8, 9, \dots 15\}$ abuts the corresponding processor $P\in\{127, 126, \dots, 120\}$.  More generally, a processor on the border with triplet $(i,j,0)$ directly borders a processor with triplet $([P_x/2 - (i+1)]\, \bmod P_x,P_y-(j+1),0)$.  As with the unshifted Klein boundaries, the set of neighboring gridpoints are again mirrored across the boundary so, for example, the leftmost gridpoint in the bottom row of processor 1 is considered to be directly above the rightmost gridpoint of processor 118.


\section{Simulation Results}\label{sec:results}
We now present results from simulations using the various types of boundary conditions discussed above. The computational domain measures $L_x\times L_y = 51.2d_i\times12.8d_i$ for the GEM Challenge and shifted Klein cases and is doubled in $L_y$ for the doubly periodic case.  (In other words, each reconnecting current sheet has the same aspect ratio in all runs.)  The reconnecting component of the magnetic field has the initial form shown in either equation \ref{eq:gem} or \ref{eq:doubleharris} while the density has the asymptotic value of $0.2n_0$ and varies as necessary to ensure force balance in the $y$ direction.  Section \ref{sec:antiparallel} considers the case where the asymptotic fields are anti-parallel, as in the original GEM Challenge, while Section \ref{sec:guide} adds a $B_z$ component that reduces the shear angle between the asymptotic fields.

\subsection{Anti-Parallel Reconnection}\label{sec:antiparallel}

\begin{figure}
\begin{center}
\includegraphics[width=\columnwidth]{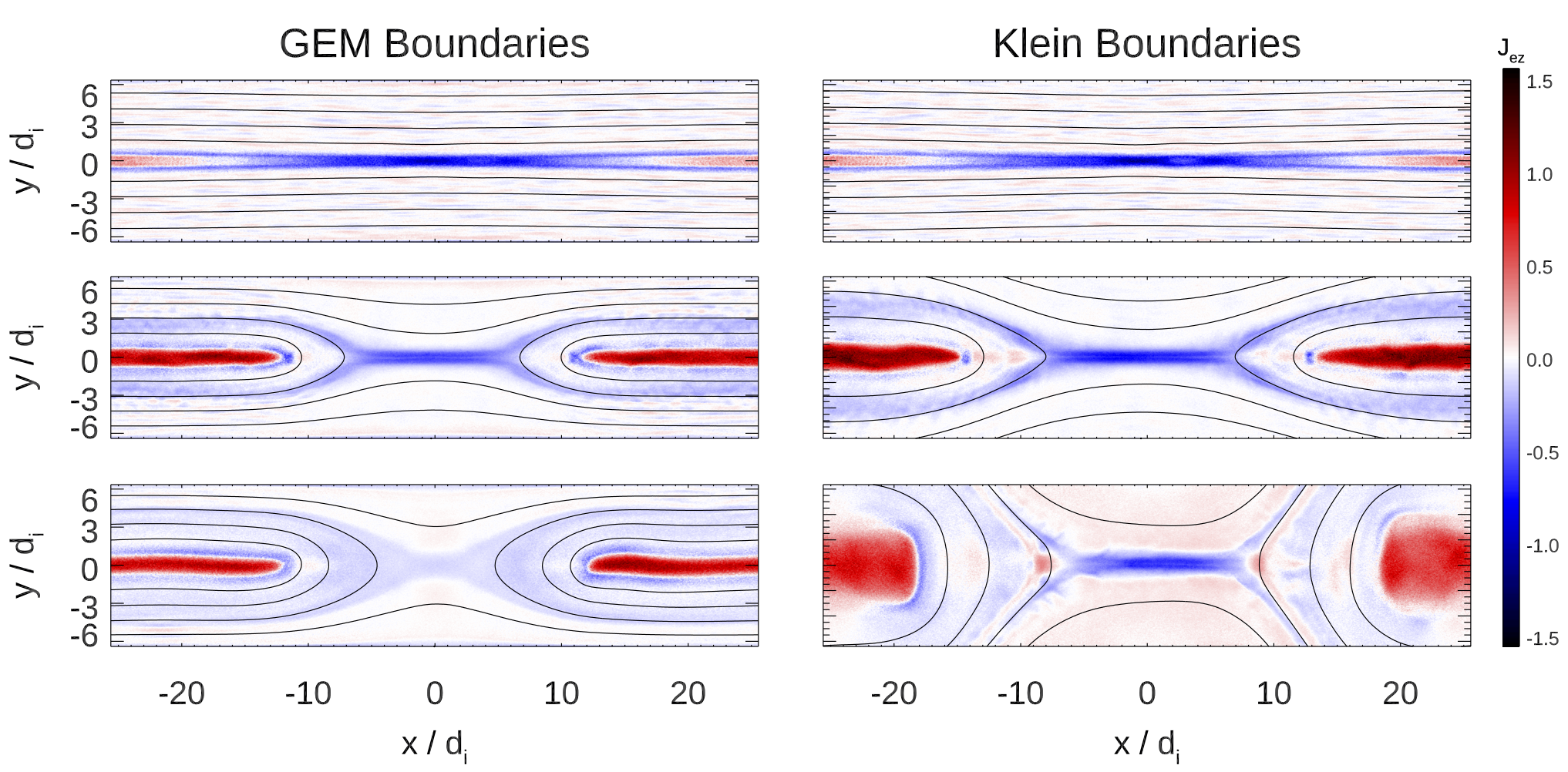}
\caption{Out-of-plane electron current density and magnetic field lines at $t\Omega_{cp} = 20$, $40$, and $60$ for the GEM Challenge boundary conditions (left) and the Klein boundary conditions (right).}\label{fig:jez_gem_klein}
\end{center}
\end{figure}

Figure \ref{fig:jez_gem_klein} shows a comparison between runs with the GEM Challenge and the shifted Klein boundary conditions with the three panels in each column depicting the simulations at $t\Omega_{cp}=20$, $40$, and $60$, respectively.  In each, the out-of-plane current density is shown in color and is overplotted with magnetic field lines.  For the GEM Challenge boundaries (left column), the islands of reconnected magnetic flux  grow freely at first (top panel) until they begin to interact with the walls at the top and bottom of the domain (middle panel).  The increasing pressure in the island exerts a growing force on plasma ejected from the X-point and suppresses further reconnection (bottom panel).  

In contrast, the panels in the right column of Figure \ref{fig:jez_gem_klein} show results from an otherwise-identical simulation that employs the shifted Klein boundary conditions described above.  By the time of the middle panel the effects of the boundary conditions are already noticeable in the magnetic field lines since, unlike in the GEM Challenge simulation, they are free to pass through the top and bottom boundaries.  By late time (bottom panel), the islands of reconnected flux have grown substantially  and, due to the shift at the boundaries, continue to drive reconnection at the X-line.  

\begin{figure}
\begin{center}
\includegraphics[width=\textwidth]{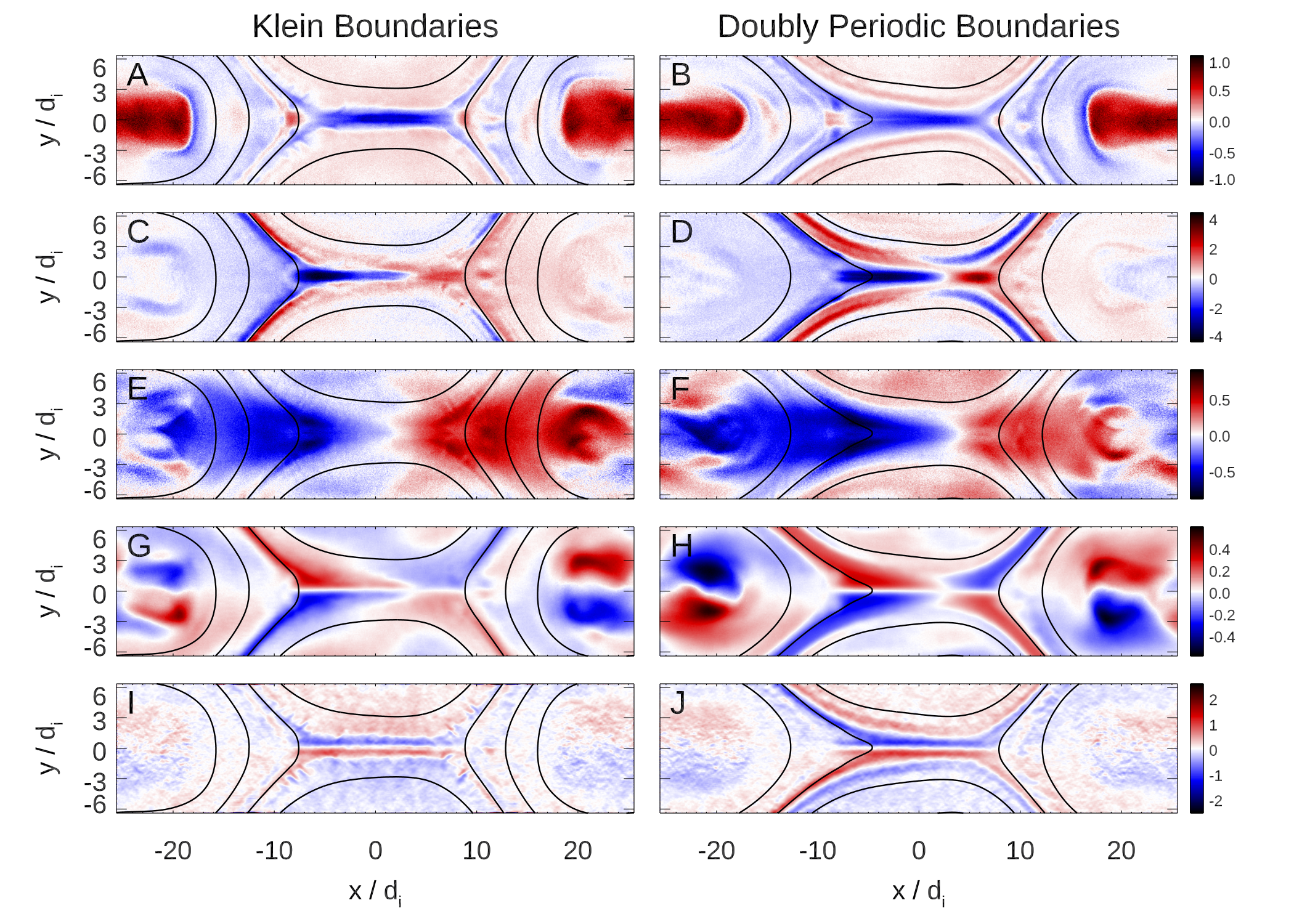}
\caption{Comparison of results with the Klein (left column) and doubly periodic (right column) boundary conditions at $t\Omega_{cp}=60$.  (A)-(B): Out-of-plane electron current density ($J_{ez}$); (C)-(D) Horizontal electron velocity ($v_{ex}$); (E)-(F): Horizontal ion velocity ($v_{ix}$); (G)-(H): $B_z$; (I)-(J): $E_y$ }\label{fig:kleindbl}
\end{center}
\end{figure}

The lack of hard walls at the top and bottom boundaries makes the Klein boundary conditions  similar to the doubly-periodic case.  Figure \ref{fig:kleindbl} presents a comparison of several quantities at $t\Omega_{cp}=60$, a time of robust reconnection, from runs with both boundary conditions.  The panels from the Klein boundaries (left column) show the entire domain, while those from the doubly periodic case (right column) show only one of the two current sheets.  The panels depict, from top to bottom, $J_{ez}$, $v_{ex}$, $v_{ix}$, $B_z$, and $E_y$.  Each horizontal pair of images uses the same color scale and so may be directly compared.  The two columns are quite similar, suggesting that the Klein boundary conditions do not significantly affect the evolution of the system.

\begin{figure}
\begin{center}
\includegraphics[width=\columnwidth]{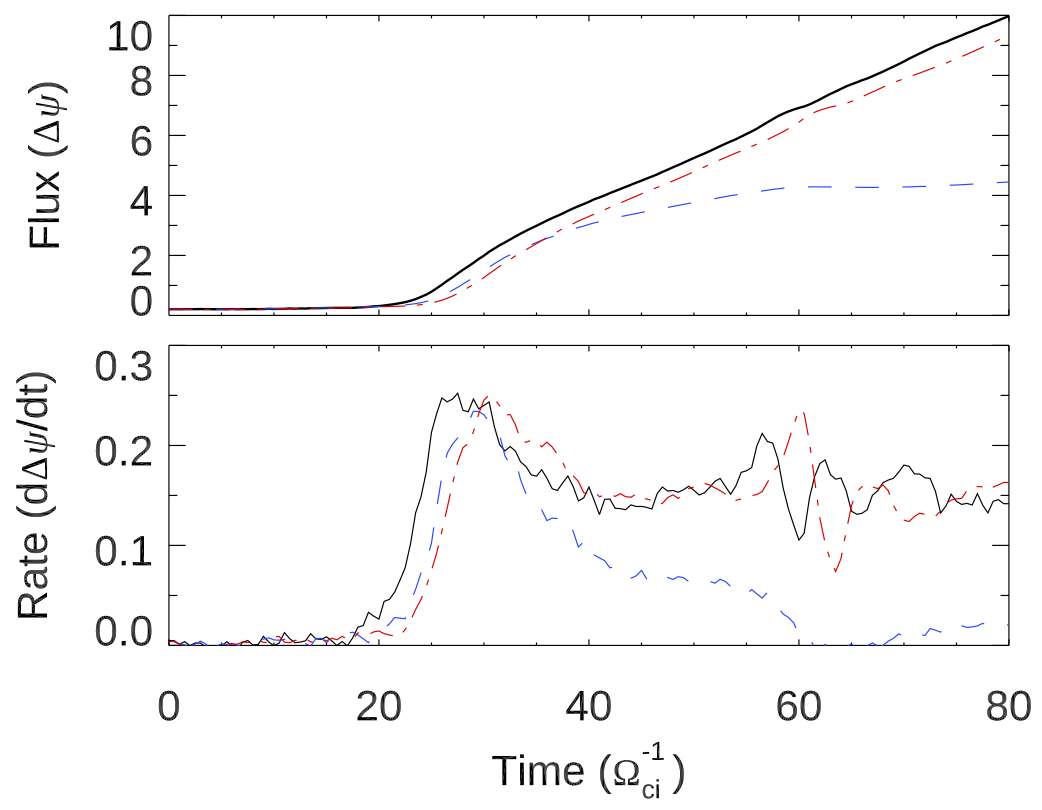}
\caption{Reconnected flux (top), $\Delta\psi$, and reconnection rate (bottom), $d(\Delta\psi)/dt$ versus time for the GEM Challenge (blue dash), doubly periodic (red dot-dash), and shifted Klein (black) boundary conditions.}\label{fig:fluxesandrates}
\end{center}
\end{figure}

We next compare all three runs in Figure \ref{fig:fluxesandrates}, which plots the reconnected flux, $\Delta\psi$, and the reconnection rate, $d(\Delta\psi)/dt$, versus time for the different cases: GEM Challenge in blue dashes, doubly periodic in red dot-dashes, and shifted Klein in black.  Contours of $\psi$ are the magnetic field lines shown in Figures \ref{fig:jez_gem_klein} and \ref{fig:kleindbl} and in the first and third cases $\Delta\psi$ is merely the difference in $\psi$ between the X- and O-points.  In the doubly periodic case the curve plots the average of the differences observed in the two current sheets.  Each of the simulations exhibit an initial overshoot where the reconnection rate spikes to $\approx 0.25$, but in the GEM Challenge case that spike is followed by a steady decline, with reconnection essentially ending after $t\Omega_{cp}\approx 50$ when $\Delta\psi$ plateaus.   In contrast, the doubly periodic and shifted Klein runs both settle into a state with reasonably constant reconnection between $t\Omega_{cp}=40$ and $t\Omega_{cp}=80$ with $d(\Delta\psi)/dt \approx 0.15$. The large excursions near $t\Omega_{cp}=60$ correspond to the time when the once-reconnected field lines surrounding the magnetic islands interact with an X-point a second time, either that on the opposite current sheet for the doubly periodic case or the original one for the shifted Klein (see Figure \ref{fig:jez_gem_klein}).  It is interesting to note that the reconnection rate after these fluctuations roughly remains unchanged.

The elapsed time (measured on a clock) for the three computations differed by less than $3\%$. However, the GEM Challenge and shifted Klein cases were run on the same number of processors, while the doubly periodic case, for which the computational domain was doubled in size, used twice as many.  Hence, while the doubly periodic and shifted Klein cases found very similar physical results, the former incurred twice the computational expense of the latter.  (For the relatively small simulations considered here, $\texttt{p3d}$ exhibits near-perfect weak scaling and so any extra overhead incurred by doubling the number of processors is minimal.)

\subsection{Guide Field Reconnection}\label{sec:guide}

Adding a spatially constant guide field, $B_g$ to the initial conditions of the GEM Challenge and doubly periodic cases is straightforward.  The Klein case is more subtle.  Recall that the $x$ and $z$ components of vectors are flipped when crossing the Klein boundary.  As a consequence, a spatially constant guide field would exhibit a discontinuous jump -- a gridpoint at the top or bottom boundary with  $B_z = B_g$ would be adjacent to one with $B_z = -B_g$.  Such a configuration would generate a large current and lead to instability.  This behavior arises because Klein bottles are non-orientable, which means there is ambiguity in the definition of "out of the page" and "into the page".  (This is perhaps most easily seen with the M\"obius strip.  If one takes an arrow pointing normal to the surface and translates it once around the strip to the starting location, the head will point in the opposite direction.)  

To sidestep this obstacle, we can slightly modify the boundary condition. Write $B_z$ as the sum of a constant background $B_{z0}$ and a perturbation: $B_z = B_{z0} + \delta B_z$.  Then, rather than apply the Klein boundary to the full magnitude, only change the sign of the perturbation $\delta B_z$ when it crosses the boundary.  Note that in the run discussed in Section \ref{sec:antiparallel}, $B_{z0}=0$ so this procedure is identical to modifying the full $B_z$.  It might be argued that, for consistency, such a modification should be applied to any $z$ component  (e.g., $E_z$ or $J_z$), but in the initial states considered here all other quantities have no uniform component and so making such an adjustment would have no effect.

\begin{figure}
\begin{center}
\includegraphics[width=\columnwidth]{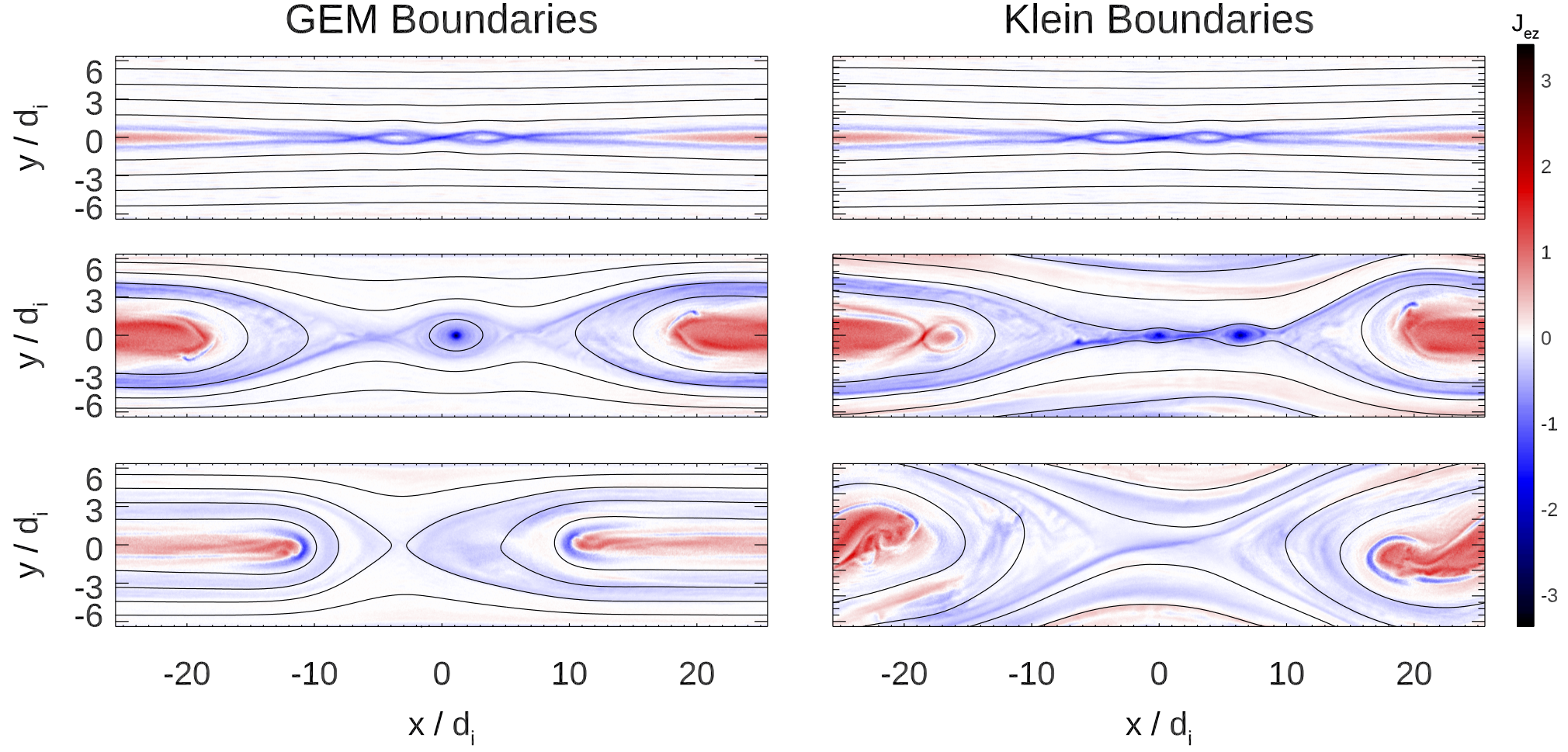}
\caption{Out-of-plane electron current density and magnetic field lines at $t\Omega_{cp} = 20$, $40$, and $60$ for the GEM Challenge boundary conditions (left) and the Klein boundary conditions (right) in runs with $B_g=1$.}\label{fig:jez_gem_kleinguide}
\end{center}
\end{figure}

We now compare results from three simulations with $B_{g0} = 1$.  Figure \ref{fig:jez_gem_kleinguide} shows the evolution of the GEM Challenge boundary conditions (left) and the Klein boundaries with the modification discussed above.  
As is typical, the addition of a guide field breaks a symmetry of the system and produces differences between the separatrices, while the reconnection layer becomes more likely to form plasmoids, as can be seen in the middle panels. As with the anti-parallel case shown in Figure \ref{fig:jez_gem_klein}, the GEM Challenge system is restricted by the walls and stagnates by the last panel.  The Klein boundaries again allow flux to pass through them as can be seen by the field lines intersecting the top and bottom of the domain.

\begin{figure}
\begin{center}
\includegraphics[width=\textwidth]{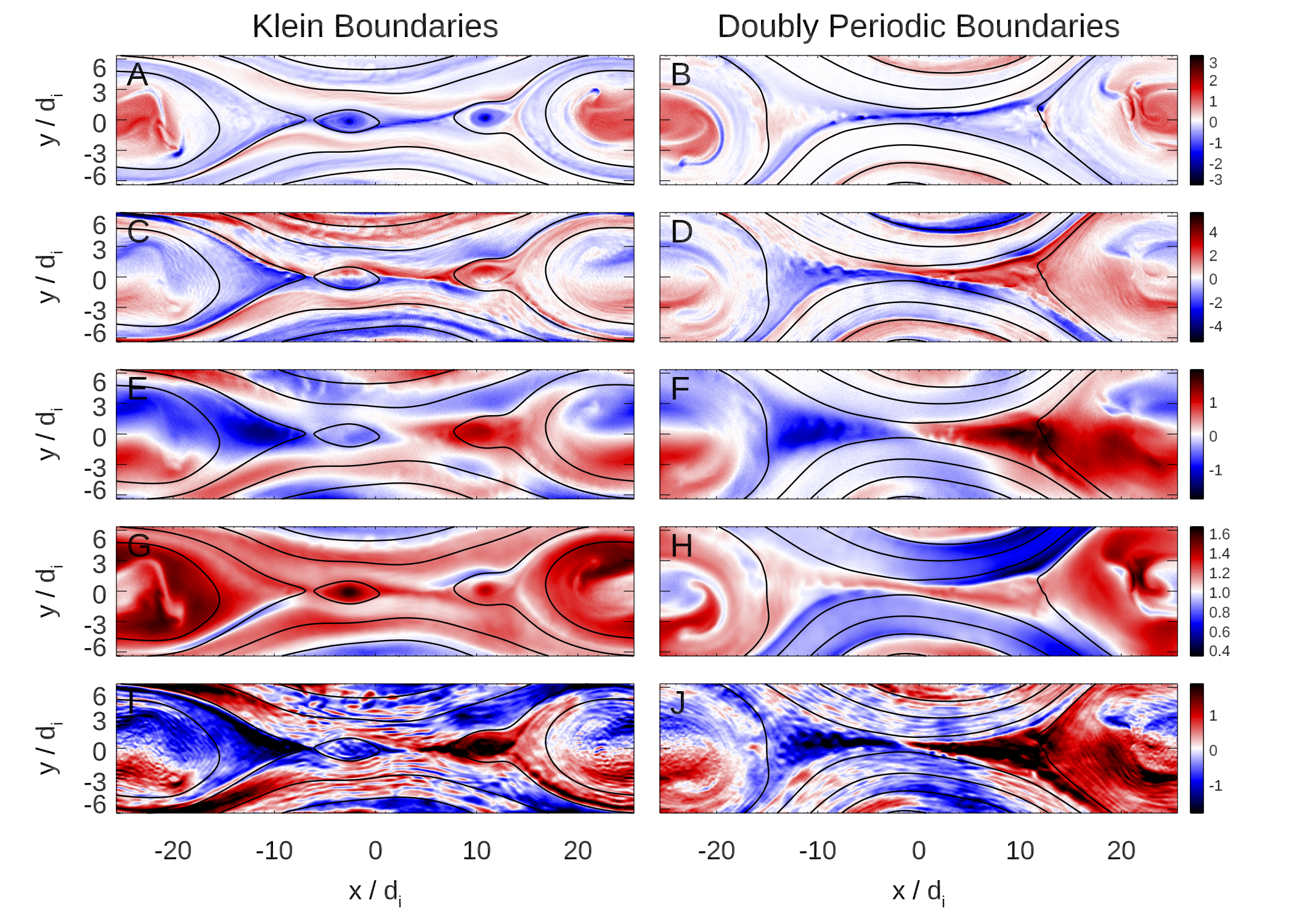}
\caption{Comparison of results with the Klein (left column) and doubly periodic (right column) boundary conditions at $t\Omega_{cp}=45$ in runs with $B_g=1$.  (A)-(B): Out-of-plane electron current density ($J_{ez}$); (C)-(D) Horizontal electron velocity ($v_{ex}$); (E)-(F): Horizontal ion velocity ($v_{ix}$); (G)-(H): $B_z$; (I)-(J): $E_y$ }\label{fig:kleinguidedbl}
\end{center}
\end{figure}

A comparison between the Klein and doubly periodic runs shows more significant differences than in the anti-parallel case.  Figure \ref{fig:kleinguidedbl} compares the two in the same format as Figure \ref{fig:kleindbl}, but at an earlier time, $t\Omega_{cp} = 45$.  The reason for this choice will be discussed below.  The plasmoids in the Klein case distort the reconnecting layer, but the overall structure is generally similar.  (Note that the color bar is the same for each pair of images so direct comparisons can be made.)  The largest differences come in panels G and H, which show $B_z$ in the two runs.  While the fields in the islands of reconnected flux are roughly comparable, bands of low-strength (blue) field are apparent just upstream of the current sheet in the doubly periodic case.  These features represent the last vestiges of unreconnected plasma being pushed towards the X-line by an island expanding on the other (unseen) sheet.  These features do not appear in the Klein boundary case, although hints are visible at the the top and bottom of the domain.  Still, the boundary conditions are clearly beginning to retard the inflow.

\begin{figure}
\begin{center}
\includegraphics[width=\columnwidth]{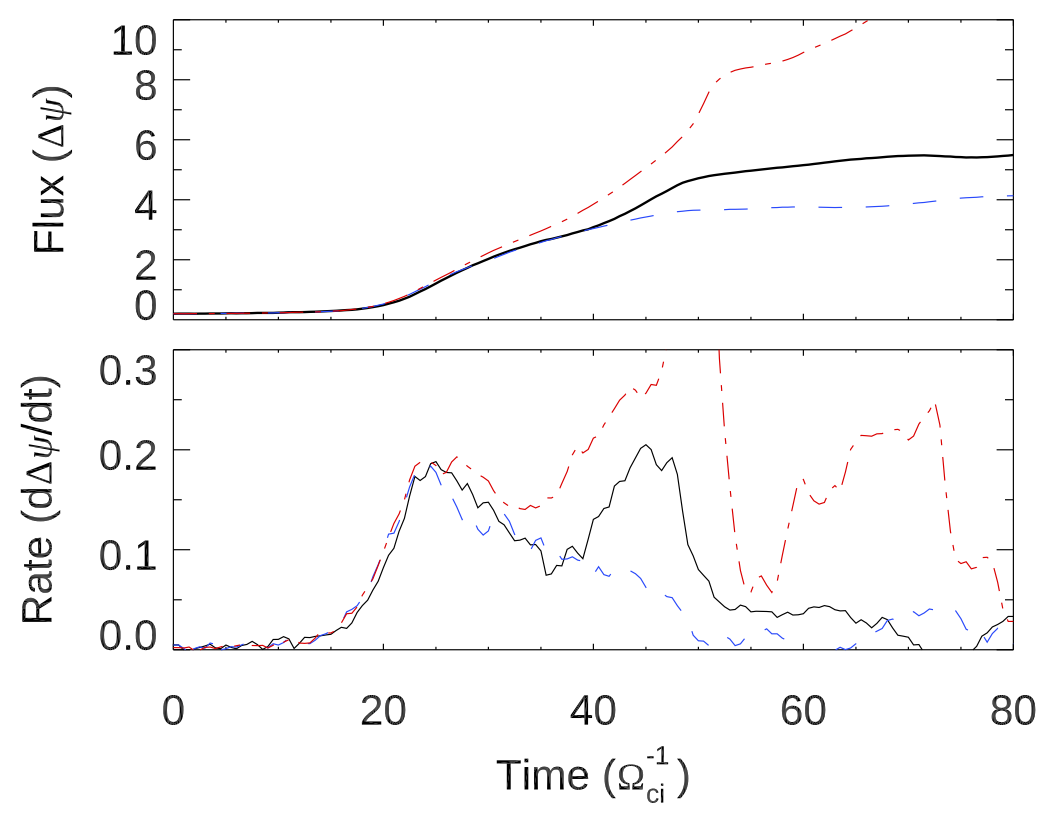}
\caption{Reconnected flux (top), $\Delta\psi$, and reconnection rate (bottom), $d(\Delta\psi)/dt$ versus time for the GEM Challenge (blue dash), doubly periodic (red dot-dash), and shifted Klein (black) boundary conditions in runs with a guide field $B_g=1$.}\label{fig:guidefluxesandrates}
\end{center}
\end{figure}

The effects of the various boundary conditions can also be seen in Figure \ref{fig:guidefluxesandrates}, which shows the reconnected fluxes and reconnection rates for the three runs in the same format as Figure \ref{fig:fluxesandrates}.  After the peak at $t\Omega_{cp} \approx 25$, the GEM Challenge run essentially stagnates, with the reconnection rate falling to zero.  The other two cases exhibit a similar earlier peak and decline, but both have a later period of increasingly fast reconnection with the Klein case peaking at $t\Omega_{cp} \approx 45$, the time shown in Figure \ref{fig:kleinguidedbl}.  After this point, however, the Klein system stagnates while the doubly periodic system continues to enjoy robust reconnection.  It is worth noting that simulations are usually halted once current sheets begin interacting with each other, so the lack of agreement after $t\Omega_{cp} \approx 50-55$ is not significant from a practical standpoint.

The elapsed times for the three computations were again similar, differing by less than $4\%$. But, as previously, the doubly periodic case incurred twice the computational expense due to its larger domain.

\section{Discussion}\label{sec:disc}
We have shown that a novel boundary condition based on the topology of a Klein bottle can be used in numerical simulations of magnetic reconnection.  The new boundary condition combines the advantages of other common options by being both self-reinforcing (as in systems with doubly periodic boundary conditions) and computationally compact (as seen in the single current sheet of the GEM Challenge boundaries).

The extension of Klein boundary conditions to full three-dimensional (3x3v) simulations should be straightforward. The additional dimension can be physically pictured as adding a "thickness" to the Klein bottle in Figure \ref{fig:kleinbottle}.  From a computational perspective, the most significant change is to the identification of neighboring processors across the Klein boundary.  Neighboring processors in $z$ must change in the manner shown in panel (ii) of Figure \ref{fig:mappings} -- recall that the $z$ component of vectors flips across the Klein boundary --  but do not have to undergo the shift of panel (iii).

However, regardless of the dimensionality, Klein boundary conditions do have limitations.  Symmetric reconnection. in which the asymptotic plasmas on the two sides of the current sheet share the same characteristics, is an idealization, albeit a common one, that sidesteps many of the complexities of actual systems.  In reality, asymmetries can exist in any or all of the plasma density, plasma temperature, strength of the reconnecting component of the field,  strength of the guide field, etc., and at some locations (e.g., the terrestrial magnetopause), reconnection is nearly always asymmetric.  Such complications are straightforward to include with GEM Challenge or doubly periodic boundary conditions because the two sides of a reconnecting current sheet represent distinct plasmas.  For Klein boundaries there is no distinction between the two sides, and so simulating asymmetric reconnection while maintaining the associated benefits (low computational cost and non-stagnation) is not straightforward.  


Finally, the topological equivalence of the doubly periodic system to a torus  implies the existence of certain conservation laws.  Specifically, consider the surface integral of Faraday's Law, $\partial \mathbf{B}/\partial t = K\boldsymbol{\nabla\times} \mathbf{E}$, where the proportionality constant $K$ depends on the choice of units.  Performing a surface integral and invoking Stokes' theorem transforms the right-hand side into a boundary integral, which vanishes because a torus has no boundary.  Hence, the magnetic flux through the plane of the simulation must be constant, which can be verified numerically.  A Klein bottle also has no boundary, but it is non-orientable and so Stokes' theorem does not apply.  As a consequence, the magnetic flux through the plane of the simulation can, and does, change when Klein boundary conditions are employed.  While this difference does not appear to affect the dynamics of reconnection, further research will be necessary to explore whether it has physical significance.



\begin{acknowledgments}
The authors acknowledge the support of NASA grants 80NSSC22K0352 and 18-DRIVE18\_2-0029, "Our Heliospheric Shield", 80NSSC22M0164).  The authors also acknowledge UMD's TREND program, funded through NSF PHY-2150399.  The authors would like to thank P. R. Colarco and E. A. Colarco for providing visual inspiration.  The simulations were carried out at
the National Energy Research Scientific Computing Center
(NERSC). The data used to perform the analysis and construct
the figures for this paper are preserved at 
\url{https://doi.org/10.5281/zenodo.12785961}.  Further data from the associated runs are stored at the NERSC High
Performance Storage System and are available upon request.
 
\end{acknowledgments}

%
\end{document}